\documentclass[aps,prx,onecolumn,groupedaddress]{revtex4}

\usepackage{amssymb,hyperref}
\usepackage{color,graphicx}
\usepackage{lmodern}
\usepackage{amsmath}
\usepackage{amsbsy}
\usepackage{float}
\usepackage{bbm}
\usepackage{bm}
\usepackage{epsfig}
\usepackage{braket}
\usepackage{caption,comment}

\newcommand{\bq}{\begin{equation}} \newcommand{\eq}{\end{equation}}
\newcommand{\bqali}{\bq\begin{aligned}}
\newcommand{\eqali}{\end{aligned}\eq}
\newcommand{\bqn}{\begin{equation*}}
\newcommand{\eqn}{\end{equation*}}
\newcommand{\TR}[2]{\text{Tr}^{(#1)}\left[#2\right]}

\newcommand\D{\operatorname{d}}
\renewcommand\k{{\bf k}}

\renewcommand\d{{\bf d}}
\renewcommand\k{{\bf k}}
\renewcommand\r{{\bf r}}
\newcommand\p{{\bf p}}
\newcommand\0{{\bf 0}}
\newcommand\E{{\bf E}}

\newcommand{\OO}{{\bm \Omega}}
\newcommand{\xii}{{\bm \xi}}
\newcommand\EXP{\operatorname{exp}}
\newcommand*{\rttensortwo}[1]{\bar{\bar{#1}}}

\newcommand\I{\operatorname{I}}

\begin{document}
\title{Rotational decoherence due to thermal photon scattering}

\author{Hamid Reza Naeij}
\email{naeij@alum.sharif.edu}
\affiliation{Azadi Ave., P.O.Box 13547‑59647, Tehran, Iran}

\date{\today}

\begin{abstract}

The use of rotational degrees of freedom of quantum systems in quantum technologies is limited by environmental effects under the decoherence mechanism. Here, we study the mechanism of decoherence based on a new formalism of elastic scattering for a non-spherical quantum system prepared in a superposition of rotational degrees of freedom. We show that for a dielectric ellipsoid immersed in an environment composed of thermal photons, the rotational decoherence rate depends on the angular differences between the two configurations of the system and the seventh power of the temperature which is different from the translational one. Then, we analyze the effect of different values of angular momentum quantum numbers of the environmental spherical harmonics on the rotational decoherence rate. 

\end{abstract}

%%\pacs{03.65.Ta, 03.65.Yz, 03.65.Nk}
\maketitle
\textbf{keywords}: Rotational decoherence, Thermal photon scattering, Spherical harmonic, Angular momentum

\section{Introduction}

The effect of the environment on the quantum systems leads to losing the quantum coherence of the system known as decoherence. Decoherence is an inevitable characteristic of any quantum system which sets the limitations in quantum technology and quantum applications. The main idea of decoherence is due to the openness of quantum systems in reality. The evolution of an open quantum system is non-unitary and follows master equations instead of Schrödinger's equation \cite{Schlosshauer,Breuer}. In recent years, there are many works in some fields of quantum mechanics such as quantum foundations, quantum optics, quantum information, quantum biology, etc that investigate the effect of the environment on quantum systems and decoherence theory; the literature on this subject is vast \cite{Salari,Naeij1,Soltanmanesh1,Naeij2,Soltanmanesh2,Tirandaz,Naeij3,Seifi,Soltanmanesh3}.

New advances in quantum technology provide new perspectives for exploiting of rotational degrees of freedom of quantum systems. In this regard, many experimental works focused on the study of rotational degrees of freedom and control of molecules \cite{Shepperson,Tenney} and nanoparticles \cite{Hoang, Kuhn1}. Moreover, the possible applications have been investigated in ultracold chemistry \cite{Moses}, highly sensitive sensors \cite{Kuhn2}, quantum heat engine \cite{Roulet} and nanomagnets \cite{Rusconi}. It seems that understanding of these experiments and applications requires the study of theoretical aspects of the effects of the environment on the rotational degrees of freedom of quantum systems. However, in the past decades, most studies have been focused on the effect of the environment on the translational degrees of freedom, e.g.; see \cite{Joos,Hornberger,Aspelmeyer,Caldeira,Becker} and, just a few studies focused on the case of rotations.

A first master equation in rotational decoherence was reported in \cite{Zhong}, and then extended in \cite{Stickler}. In our previous work in rotational decoherence \cite{Carlesso} we presented a general perturbative approach to compute the decoherence rate for a quantum system in a superposition of its rotational degrees of freedom. The specific cases of the dipole-dipole and quadrupole-quadrupole interactions are solved explicitly, and we show that the rotational degrees of freedom decohere on a time scale that can be longer than the translational one.

In this paper, we want to dig further into the characteristics of the decoherence mechanism for a quantum system prepared in a superposition of rotational degrees of freedom based on a model of elastic scattering.  As we know, one of the main sources of decoherence and quantum to classical transition is the scattering with environmental particles. Here, we first obtain a general and exact expression for the decoherence rate. Then, we investigate the details of the rotational decoherence of a quantum system in the presence of an environment composed of thermal photons. Our results show that for a dielectric ellipsoid immersed in an environment composed of thermal photons, assuming no absorption, the rotational decoherence rate depends on the difference of scattering amplitudes for different rotational orientations and can be shown to depend on the difference in the angles of orientations. Furthermore, the decoherence rate is proportional to the seventh power of the temperature which is different from the translational one. Then, we analyze the effect of different values of angular momentum quantum numbers of the environmental spherical harmonics on rotational decoherence rate.

In the sections that follow, we first introduce the new formalism to derive the master equation and rotational decoherence rate. Then, we calculate the decoherence rate of the system prepared in a superposition of rotational degrees due to thermal photon scattering. Moreover, we analyze the effect of spherical environment characteristics on the rotational decoherence rate. Finally, the results are discussed in the conclusion section.

\section{Formalism}

The system is modeled as a non-spherical particle of mass $M$, whose orientation can be described in terms of the states $\ket \OO$, representing the system prepared in the angular configuration $\OO$. Such a configuration can be prepared by starting from some reference configuration $\ket{{\bm 0}}$ (e.g., in a way that the anisotropy of the system is aligned along the $x$ axis) and applying a rotation $\hat D_\text{\tiny S}(\OO)=e^{-i\hat L_z\gamma/\hbar}e^{-i\hat L_x\beta/\hbar}e^{-i\hat L_z\alpha/\hbar}$ defined by the three Euler angles $ \OO=({\alpha,\beta,\gamma})$ \cite{Sakurai}. 

The statistical operator describing the rotational state of the system is
\bq
\hat \rho_\text{\tiny S}=\int \D\OO\int \D\OO'\,{\rho}_\text{\tiny S}(\OO,\OO')\ket \OO \bra{\OO'},
\eq
where $\rho_\text{\tiny S}(\OO, \OO')$ is its matrix elements with respect to $\ket{\OO}$ and $\ket{\OO'}$. 
At time $t=0$, we consider the system as decoupled from its environment, which is assumed to be in thermal equilibrium. Thus, the total initial state is $\hat \rho_\text{\tiny T}=\hat \rho_\text{\tiny S}\otimes\hat \rho_\text{\tiny E}$, where $\hat \rho_\text{\tiny E}$ is the state of the environment.

Considering  the configuration ${\bm\Omega}$, a scattering process at time $t$ can be written as follows
\bq
\ket{\OO}\otimes\ket{\chi}		\quad\xrightarrow[]{\text{scattering}}\quad		\ket{ \OO}\otimes\hat{\mathbb S}_{\OO}\ket{\chi},
\eq
where the recoil-less limit is considered ($M\gg m_E$, where $m_E$ is the mass of the environmental particles) and thus the scattering operator $\hat{\mathbb S}_{\bm \Omega}$ acts on the environmental state $\ket\chi$ only. $\hat{\mathbb S}_{\bm \Omega}$ can be related to the standard unitary scattering operator $\hat{\mathbb S}_{0}$ acting in  ${\bm \Omega}={\bm 0}$, through a rotation from the configuration $\ket{\bm 0}$ to $\ket{\bm \Omega}$. So we have $\hat{\mathbb S}_{\bm \Omega}=\hat D_\text{\tiny E}({\bm \Omega})\hat{\mathbb S}_{0}\hat D_\text{\tiny E}^\dag({\bm \Omega})$.

At time $t$, after the scattering process, the reduced state of the system is given by
\bq
\hat \rho_\text{\tiny S}(t)=\int \D{\bm \Omega}\int \D{\bm \Omega'}\,{\rho}_\text{\tiny S}(\OO,\OO') \ket{{\bm \Omega}}\bra{{\bm \Omega}'}\eta({\bm \Omega},{\bm \Omega}'),
\eq
where
\bq
\eta({\bm \Omega},{\bm \Omega}')=\TR{E}{\hat\rho_\text{\tiny E}\hat{\mathbb S}_{{\bm \Omega}'}^{\dag}\hat{\mathbb S}_{{\bm \Omega}}},
\eq

To explicitly calculate the partial trace over the degrees of freedom of the environmental particles, we consider the total system to be confined in a box of volume $V$ , and we assume the environment is at thermal equilibrium. Therefore, the state of the environment can be written as

\bqali
\hat \rho_\text{\tiny E}=\frac{(2\pi)^3}{V}\int\D k\, k^2\mu(k)\sum_{l=0}^{+\infty}\sum_{m=-l}^l \vert k,l,m\rangle \langle k,l,m \vert ,
\eqali

Here, we assumed that the momentum distribution of the environmental particles is invariant under rotations, thus $\mu(\k)=\mu(k)$. Moreover, $\ket{k,l,m}$ is the common eigenstate of the momentum $\hat{P}^2$, the total angular momentum $\hat{L}^2$, and its $z$ component $\hat{L}_z$ of the environmental particles. The relation between the usual momentum eigenstate $\ket{\p}$ and $\ket{k,l,m}$ can be written as \cite{Sakurai}

\bq
\braket{\p|k,l,m}=\frac{\delta(p-k)}{ p}Y_{l,m}(\hat \p),
\eq
where $Y_{l,m}(\hat \p)$ denotes the spherical harmonic and $\hat \p=\p/p$.

Let us consider the case in which the system is in a superposition of angular configuration around the $z$ axis. This case can be interesting in the experimental setup where one usually considers one direction per time. So, the state of the system can be described as  $\ket \alpha=\hat{D}_E(\alpha)\ket \0$ where $\hat{D}_E(\alpha)=\EXP(-\frac{i}{\hbar}\hat{L}_z\alpha)$. So, we have

\bqali
\eta(\alpha,\alpha')=\frac{(2\pi)^3}{V}\int\D k\, k^2\mu(k)\sum_{l=0}^{+\infty}\sum_{m=-l}^l\braket{k,l,m|\hat{\mathbb S}_{{\alpha}'}^{\dag}\hat{\mathbb S}_{{\alpha}}|k,l,m},
\eqali
where $\hat{D}_E(\alpha)\ket{k,l,m}=e^{-im\alpha}\ket{k,l,m}$.

The scattering matrix can be defined as 
\bqali
\hat{\mathbb S}_{{\alpha}}=\hat{D}_E(\alpha)(1+i\hat{T}) \hat{D}^\dag_E(\alpha),
\eqali
where $\hat{T}$ is the $T$ matrix in scattering theory \cite{Sakurai}. Due to the unitary nature of $\hat{\mathbb S}_{{\alpha}}$, one can show $-i(\hat{T}^\dag-\hat{T})=\hat{T}^\dag\hat{T}$. By using this relation we find

\begin{widetext}
\bq
\eta({\alpha},{ \alpha}')=1-\frac{(2\pi)^3}{V}\int\D k\, k^2\mu(k)\sum_{lm}\int\D p\,p^2 \sum_{l'm'}\left(1-e^{-i(m-m')(\alpha-\alpha')}\right)\braket{k,l,m|\hat T^\dag|p,l',m'}\braket{p,l',m'|\hat T|k,l,m},
\eq
\end{widetext}
 The matrix elements of $\hat T$ can be written in the momentum space as \cite{Sakurai}
\bq
\braket{\k''|\hat T|\p''}%=-\frac{\hbar^2}{2\pi M}\delta\left[\tfrac{\hbar^2}{2M}\left((\k'')^2-(\p'')^2\right)\right]f(\k'',\p''),\\
=-\frac{1}{2\pi p''}\delta(k''-p'')f(\k'',\p''),
\eq
where $f(\k'',\p'')$ is the scattering amplitude. 

Regarding Eqs. (6) and (10), we have
\bqali
\braket{p,l',m'|\hat T|k,l,m}=-\int\D\hat \k''\int\D\p''Y_{l,m}(\hat\k'')Y^*_{l',m'}(\hat \p'')f(k\hat\k'',k\hat\p'')\frac{\delta(p-k)}{2\pi},
\eqali

Thus, we obtain
\begin{widetext}
\bqali
\eta({\alpha},{ \alpha}')=&1-\frac{t}{V}\int\D k\,\frac{\hbar k}{m_E}k^2 \mu(k)\sum_{lm}\sum_{l'm'}\left(1-e^{-i(m-m')(\alpha-\alpha')}\right)\\
&\int\D\hat\k'\int\D\hat\p'\int \D\hat\k''\int\D\hat\p''\,Y_{l,m}(\hat \k'')Y_{l,m}^*(\hat \k')
Y_{l',m'}(\hat \p')Y_{l',m'}^*(\hat \p'')f^*(k\hat\k',k\hat\p')f(k\hat\k'',k\hat\p''),
\eqali
\end{widetext}
where we exploited the well-known normalization of the squared Dirac-$\delta$:
\bq
\left(\delta(p-k)\right)^2\sim\frac{\hbar p t}{2\pi m_E}\delta(p-k),
\eq
which is valid under the assumption that the decoherence time is larger than the time of the collision \cite{Schlosshauer}.

We now consider that the spherical harmonics are of the form
\bq
Y_{l,m}(\hat\k)=(-1)^m \sqrt {\frac{(2l+1)(l-m)!}{2\pi(l+m)!}}F_{l,m}(\theta_\k)e^{im\phi_\k},
\eq
where $(\theta_\k,\phi_\k)$ describe $\hat \k$ and $F_{lm}(\theta)$ is the Legendre polynomial function. Therefore, we can rewrite

\bq
Y_{l,m}(\hat\k'')e^{-im\alpha}=Y_{l,m}(\hat\k''_{\alpha}),
\eq
where $\hat\k''_{\alpha}$ is obtained from $\hat\k''$ after a rotation of $\alpha$ around $z$. Consequently, the phase terms in Eq. (12) can be inserted in the spherical harmonics, and we obtain

\bqali
e^{-i(m-m')(\alpha-\alpha')}Y_{l,m}(\hat \k'')Y_{l,m}^*(\hat \k') Y_{l',m'}(\hat \p')Y_{l',m'}^*(\hat \p'') =Y_{l,m}(\hat \k''_\alpha)Y_{l,m}^*(\hat \k'_{\alpha'}) Y_{l',m'}(\hat \p'_{\alpha'})Y_{l',m'}^*(\hat \p''_\alpha),
\eqali

Considering the orthonormality of the spherical harmonics as \cite{Zettili}
\bq
\sum_{lm}Y_{l,m}(\hat \k)Y^*_{l,m}(\hat \k')=\delta(\hat \k-\hat \k'),
\eq
we obtain
\begin{widetext}
\bq
\eta(\alpha,\alpha')=1-\frac{t}{V}\int\D k\, \frac{\hbar k}{m_E}k^2 \mu(k)\int\D\hat\k'\int \D \hat\p'f(k\hat \k',k \hat \p')\left(f^*(k\hat \k',k \hat \p')-f^*(k\hat \k'_\omega,k \hat \p'_\omega)\right),
\eq
\end{widetext}
where $\omega=\alpha-\alpha'$ is the difference in the angles of the rotational configurations. From the latter expression, the system evolution can be described by the following master equation
\bq
\frac{\D \rho_\text{\tiny S}(\alpha,\alpha',t)}{\D t}=-\Lambda_R\rho_\text{\tiny S}(\alpha,\alpha',t),
\eq
where $\Lambda_R$ is the decoherence rate due to the $N$ particles of which the environment is made and defined as
\bq\label{expr5}
\Lambda_R=n\!\int\D k\, v(k) \rho(k)\frac{\int\D\hat\k'\int \D \hat\p'}{8\pi}|\Delta f^\omega(k\hat\k',k\hat\p')|^2,
\eq

Here we defined $n=N/V$ as the number density of the environment, $v(k)=\hbar k/m_E$ is the environmental particle velocity, $\rho(k)=4\pi k^2 \mu(k)$ is the momentum distribution environmental particles, and
\bq
\Delta f^\omega(\k',\p')=f(\k',\p')-f(\k'_\omega,\p'_\omega),
\eq
where $f(\k'_\omega,\p'_\omega)$ has the same form of $f(\k',\p')$ but with $\k'$ and $\p'$ replaced by $\k'_\omega$ and $\p'_\omega$, respectively, which are the same vectors rotated by the angle $\omega$. A derivation of the master equation in Eq. (19) alternative to that in \cite{Zhong} and is reported in our previous work \cite{Carlesso}.

\section{Thermal Photon Scattering}
\subsection{Calculation of $\Lambda_R$}

To further analyze the properties of the rotational decoherence rate $\Lambda_R$, we model the environmental particles as thermal photons. As we know, thermal photon scattering is a common source of decoherence of quantum states.

We consider the system is a dielectric ellipsoid immersed in an environment composed of thermal photons. We assume that $N$ photons in volume $V$ have a Planck probability distribution as
\bq
\mu(k)=\frac{V}{N}\frac{2}{\exp(\frac{\hbar ck}{k_B T})-1},
\eq
where $c$ is the speed of light, $T$ is blackbody radiation temperature and factor 2 is due to the fact that there are two possible polarization directions of the photon. The scattering amplitude for a dielectric object is obtained by the scattered radiation from the induced dipole. For a dielectric, the cross-section is obtained by the scattered radiation from the induced dipole. For an incoming field $\E_{inc}$ with a polarization vector $\xii$, the scattering amplitude can be defined as \cite{Jackson}
\bq
f(\k',\p')=\frac{k^2}{4\pi\varepsilon_0 E}\xii'.\d ,
\eq
where $\xii'$ is the polarization vector of the outgoing radiation and $\d$ is the induced dipole moment. The induced dipole moment is defined as $\d=\rttensortwo{\alpha}_\OO.\E_{inc}$ where $\rttensortwo{\alpha}_\OO$ is the polarizability of the ellipsoid with configuration $\OO$ and the incoming field is $\E_{inc}=\xii E \exp(-i \k'.\r)$. Moreover, we assume the polarizability is diagonal
\bq
\rttensortwo{\alpha}_0=
\begin{pmatrix}
\alpha_x & 0 &0 \\
0 & \alpha_y & 0 \\
0 & 0 & \alpha_z 
\end{pmatrix},
\eq
where the subscript means the Euler angles are zero. The polarizability of the ellipsoid in our model can be obtained as
\bq
 \rttensortwo{\alpha}_{\omega}=\hat{R}_z^\dagger(\omega)\rttensortwo{\alpha}_0 \hat{R}_z(\omega) 
 \eq
where $\hat{R}_z(\omega)$ is rotation matrix around $z$ axis.

We now calculate the difference of the scattering amplitudes in Eq. (20) using Eq. (23). We obtain

\bqali
\Lambda_R=n\!\int\D k\, v(k) \rho(k)\frac{\int\D\hat\k'\int \D \hat\p'}{8\pi} \frac{k^4}{(4\pi \epsilon_0)^2}|\xii'.(\rttensortwo{\alpha}_0-\rttensortwo{\alpha}_{\omega}). \xii)|^2,
\eqali

Consequently, using the following identity
\bq
\sum_\lambda \xi^{(\lambda)}_i \xi^{(\lambda)}_j=\delta_{ij}-\hat{\k}_i.\hat{\k}_j ,
\eq
where $\lambda$ is the polarization index, and Eq. (22) for the probability distribution function, we obtain the decoherence rate as
\bq
\Lambda= 6!\frac{ 2c }{9\varepsilon_0 ^2 } (\frac{k_B T}{\hbar c})^7 \zeta (7) (\alpha_x-\alpha_y)^2 \sin^2\omega ,
\eq
where $\zeta (7)=\simeq 1.00835$ and $c=\hbar k/m_E$.

The above equation shows that the decoherence rate of a dielectric ellipsoid depends on the angular differences between the two configurations of the system, $\omega$, and the polarizability components $\alpha_x, \alpha_y$. Moreover, the rotational decoherence rate is proportional to the seventh power of the temperature which is different from the translational one. Referring to the translational case, the decoherence rate depends on the ninth power of the temperature \cite{Joos}. The rotational decoherence rate in Eq. (28) is similar to the results reported in \cite{Zhong}.

In the next section, we will analyze the effect of different values of angular momentum quantum numbers of the environmental spherical harmonics $l, l', m, m'$ on rotational decoherence rate $\Lambda_R$.

\subsection{Effect of angular momentum quantum numbers of the environmental spherical harmonics on $\Lambda_R$}

First, we can rewrite the rotational decoherence rate $\Lambda_R$ regarding Eqs. (12)  and (16) as
\bq
\Lambda=\sum_{l,l'=0}^{\infty}\Lambda_{ll'},
\eq
where
\bq
\Lambda_{ll'}=\frac{c N }{V }\int \D k k^2 \mu (k)\big(\I_{ll'}(0)-\I_{ll'}(\omega)\big)  ,
\eq
We define
\bqali
\I_{ll'}(\omega) =\sum_{m,m'=-l}^{l} \int \D\hat{\k}'_{\alpha'} \int \D\hat{\p}'_{\alpha'} \int \D\hat{\k}''_{\alpha}\int \D\hat{\p}'' _{\alpha}Y_{lm}(\hat{\k}''_{\alpha}) Y^*_{lm}(\hat{\k}'_{\alpha'})Y_{l'm'}(\hat{\p}'_{\alpha'}) Y^*_{l'm'}(\hat{\p}''_{\alpha})f^*(\k'_{\alpha'}, \p'_{\alpha'}) f(\k''_{\alpha},\p''_{\alpha}),
\eqali
in which we used the parameters $\k'_{\alpha'}, \k''_{\alpha}, \p'_{\alpha'}, \p''_{\alpha} $ with a shift in the definition of $\k', \k'', \p', \p''$ of a quantity $\alpha$ and $\alpha'$ for the environmental spherical harmonics considering Eq. (15).

We use the addition theorem for spherical harmonics as \cite{Hassani}
\bq
\sum_{m=-l}^{l} Y_{lm}(\hat{\k}) Y^*_{lm}(\hat{\k}')=\frac{2l+1}{4\pi} P_l(\cos\gamma),
\eq
where $P_l(\cos\gamma)$ is the Legendre polynomials and $\gamma$ is the angle between $\hat{\k}$ and $\hat{\k}'$. So, we have
\bqali
\I_{ll'}(\omega)& =\frac{(2l+1)(2l'+1)}{(4\pi) ^2}
\int \D\hat{\k'}\int \D\hat{\p}'\int \D\hat{\k}''\int \D\hat{\p}''  P_l (\cos\gamma') P_{l'}(\cos\gamma) f^* (\k', \p') f(\k'', \p''),
\eqali
where $\gamma$ and $\gamma'$ defined as
\bqali
\cos\gamma' =\hat{\k}'_{\alpha'}.\hat{\k}''_{\alpha}=\cos \theta_{k'} \cos\theta_{k''}+\sin\theta_{k'}\sin\theta_{k''}\cos(\phi_{k'}-\phi_{k''}+\omega),
\eqali
and
\bqali
\cos\gamma=\hat{\p}'_{\alpha'}.\hat{\p}''_{\alpha}=\cos \theta_{p'} \cos\theta_{p''}+\sin\theta_{p'}\sin\theta_{p''}\cos(\phi_{p'}-\phi_{p''}+\omega),
\eqali

According to Eq.(29), we have

\bqali
\Lambda=\Lambda_{00}+(\Lambda_{01}+\Lambda_{10}+\Lambda_{11})+(\Lambda_{02}+\Lambda_{20}+\Lambda_{12}+\Lambda_{21}+\Lambda_{22}) +...,
\eqali

One can show that independent of the scattering amplitude form, $\Lambda_{00}$ is equal to zero, since $\I_{00}(\omega)=\I_{00}(0)$. We now analyze the others term in Eq. (36).

Considering the scattering amplitude form of thermal photons in Eq. (23), we calculate $\I_{ll'}(\omega)$ in Eq. (31) for the different values of $l$ and $l'$ such as $\lbrace0,1,2,3\rbrace$. Our calculations show that only in the case of $l=l'=1$, $I_{ll'} (\omega)$ is not equal to zero and can be obtained as
\bq
\I_{11}(\omega)=\frac{k^4}{9\epsilon_0 ^2}\big(-\sin^2 \omega(\alpha_x-\alpha_y)^2+\alpha_x^2+\alpha_y^2+\alpha_z^2\big),
\eq

Inserting Eq. (37) into Eq. (30) and using Planck probability distribution function, we obtain 
\bq
\Lambda_{11} =6!\frac{2c}{9\varepsilon_0 ^2} (\frac{k_B T}{\hbar c})^7 \zeta (7) (\alpha_x-\alpha_y)^2 \sin^2\omega ,
\eq

As is clear the rotational decoherence rate $\Lambda_R$ due to thermal photon scattering in Eq. (28) and the rotational decoherence rate in the case of $l=l'=1$ for environmental spherical harmonics in Eq. (38) are equal.

\section{Conclusion}

In this paper, we investigated the decoherence of a quantum system prepared in a superposition of rotational degrees of freedom based on a model of elastic scattering.  First, we reviewed a general and exact expression for the rotational decoherence rate based on the master equation approach which we proposed in our previous work. Then, we analyzed the details of the rotational decoherence of a quantum system in the presence of an environment composed of thermal photons (photon-gas environment).  Our results show that for a dielectric ellipsoid immersed in such an environment, assuming no absorption, the rotational decoherence rate depends on the angular differences between the two configurations of the system. Moreover, the decoherence rate is proportional to the seventh power of the temperature which is different from the translational one. Then, we analyzed the effect of different values of angular momentum quantum numbers of the environmental spherical harmonics $l, m$ on rotational decoherence rate. Our results show that the rotational decoherence rate due to thermal photon scattering and the rotational decoherence rate in the case of $l=1$ for the environmental spherical harmonics are equal. In other words, the only angular momentum quantum number affecting the rotational decoherence rate is related to the case of $l=1$.

This present work can help to better understand the role of decoherence of rotatinal degrees of freedom in experimental studies, especially in manufacturing of quantum devices.

\end{document}